# Machine learning prediction of cancer cell sensitivity to drugs based on genomic and chemical properties

Michael P. Menden[1], Francesco Iorio[1,2], Mathew Garnett[2], Ultan McDermott[2], Cyril H. Benes[3], Pedro J. Ballester[1,*], Julio Saez-Rodriguez[1,*]

[1] EMBL – European Bioinformatics Institute, Wellcome Trust Genome Campus -- Cambridge CB10 1SD, UK

[2] Cancer Genome Project, Wellcome Trust Sanger Institute, Wellcome Trust Genome Campus -- Cambridge CB10 1SD, UK

[3] Center for Molecular Therapeutics, Massachusetts General Hospital Cancer Center and Harvard Medical School, 149 13th Street, Charlestown, MA 02129, USA

* to whom correspondence should be addressed (pedro.ballester@ebi.ac.uk, saezrodriguez@ebi.ac.uk)

**Abstract**

Predicting the response of a specific cancer to a therapy is a major goal in modern oncology that should ultimately lead to a personalised treatment. High-throughput screenings of potentially active compounds against a panel of genomically heterogeneous cancer cell lines have unveiled multiple relationships between genomic alterations and drug responses. Various computational approaches have been proposed to predict sensitivity based on genomic features, while others have used the chemical properties of the drugs to ascertain their effect. In an effort to integrate these complementary approaches, we developed machine learning models to predict the response of cancer cell lines to drug treatment, quantified through $IC_{50}$ values, based on both the genomic features of the cell lines and the chemical properties of the considered drugs. Models predicted $IC_{50}$ values in a 8-fold cross-validation and an independent *blind* test with coefficient of determination $R^2$ of 0.72 and 0.64 respectively. Furthermore, models were able to predict with comparable accuracy ($R^2$ of 0.61) IC50s of cell lines from a tissue not used in the training stage. Our *in silico* models can be used to optimise the experimental design of drug-cell screenings by estimating a large proportion of missing $IC_{50}$ values rather than experimentally measure them. The implications of our results go beyond *virtual* drug screening design: potentially thousands of drugs could be probed *in silico* to systematically test their potential efficacy as anti-tumour agents based on their structure, thus providing a computational framework to identify new drug repositioning opportunities as well as ultimately be useful for personalized medicine by linking the genomic traits of patients to drug sensitivity.

**Introduction**

High-throughput screening of a large number of molecules is a widely used approach to identify lead compounds exerting a beneficial effect on a given phenotype. In the context of cancer, libraries of chemical entities have been tested in this way against panels of cell lines grown in different conditions and with heterogeneous genomic backgrounds [1]. Following the pioneering work of the "NCI-60", a collection of 59 human cancer cell lines developed by the National Cancer Institute for *in vitro* drug screening [2], recent hallmark studies have shown that screening very large cell line collections can recapitulate known and identify novel molecular genomic determinants of drug sensitivity [1,3-5].

In these studies, using systematic statistical inference and regression methods, determinant such as oncogenic lesions, high or low levels of basal gene expression and other genotypic traits have been associated to profiles of increased sensitivity/resistance to specific compounds. For instance, by applying a multivariate analysis of variance [6] and the 'Elastic Net' regression framework [7] established drug-genotype associations have been confirmed and complemented with markers of tissue-specificity and novel connections, e.g. the *EWS-FLI1* translocation in Ewing's sarcoma and sensitivity to *PARP* inhibitors, have been identified and further experimentally validated. Results of these studies have been made publicly available, providing unique resources that support the discovery of new predictive biomarkers for personalised cancer therapy.

Increasing further the size of the considered cell-line/compound panels would be very beneficial, as it provides the basis to improve the accuracy and predictive power of the inferred associations. However, this requires larger infrastructures and the cost grows with the screening size. In addition, due to various technical and logistical reasons in a high-throughput screen [7], the resulting compound-by-cell line matrix of drug efficacy (typically summarised in their $IC_{50}$, the half maximal (50%) inhibitory concentration of a substance with respect to cell viability) is often not complete. Although many steps are automated, filling experimentally each gap could be expensive and laborious [6]. Hence, an accurate tool to impute missing $IC_{50}$s and estimate them for novel cell lines would be of great value for drug screening design. Furthermore, a robust prediction tool for *in silico* identification of potentially effective drugs for treating a specific cancer could be used for drug repositioning [8,9]. An approach of this kind is represented by the COMPARE algorithm [10,11] that uses drug response profiles of the NCI-60 screening, through a 'guilt-by-association' paradigm. Following this principle, drugs eliciting a similar drug-response profile across the cell lines in the NCI-60 panel are hypothesized to share a common mode of action (MoA), thus enabling MoA discovery for novel drugs (if their tumour-suppression profile is similar to that of a known and well characterized drug) as well as the discovery of novel or secondary effects for established drugs.

Ultimately, *in silico* methods to accurately predict the effectiveness of drugs based on the molecular making of tumours (i.e. genome, transcriptome) would be a major milestone towards personalized therapies for cancer patients based on molecular biomarkers [12].

**Results**

We therefore investigated whether it is possible to build machine learning models (for details see "Materials and Methods" section, "Machine learning" subsection) that can predict drug sensitivity using cell line screening experimental data, where cell lines are treated with variable concentration of a given drug and the resulting dose-response curve summarized by an $IC_{50}$. We focused on the most comprehensive cancer drug screening dataset available to date, from the "Genomics of Drug Sensitivity in Cancer" (GDSC) project [3]. For each drug, a neural network model was trained to predict its $IC_{50}$ profile across the panel of cell lines based on the genomic background of each cell, as characterised by microsatellite instability status (1 = unstable or 0 = stable), somatic coding variants in the coding sequence of 77 cancer genes (1 = any change in protein sequence and 0 = wild type) and copy number alterations denoting gene amplification and deletion of those cancer genes (1 = amplification / more than 7 copy numbers, 0 = wild type / between 1 or 7 copy numbers, and -1 = deletion / no copy number). However, the predictive power of these initial models was limited, especially for those drugs without a well-known oncogene-to-drug response dependency.

We reasoned that cancer cell sensitivity to drug molecules is driven by features from both cells and drugs. Whereas cell features are ultimately connected to the inner workings of the cell, drug features include physicochemical properties that are correlated with the ability of the molecule to cross the cell membrane (e.g. lipophilicity) or its selectivity to intracellular targets (e.g. fingerprints encoding the chemical structure).
Indeed, extensive work has been done on Quantitative Structure-Activity Relationship (QSAR) approaches to predicting whole-cell activity of molecules based of their chemical properties [13-16], including applications to predicting anti-cancer activity in drugs [17,18]. However, such QSAR approaches exclusively based on chemical features cannot distinguish between resistant and sensitive cell lines. For instance, building a model without any information of the cell lines, the model will be not capable of predicting cell line A to be more resistant than cell line B to drug C, which is the main aim of integrating chemical and genomic features in our models.

We therefore extended our machine learning models to include as input chemical features from the drugs, besides the molecular characterization of the cell lines (see Fig 1). This integrative approach not only integrates two complementary streams of information, but also allows the model to be trained with much larger amounts of data, which is often a key factor to improve predictive performance (see Fig 2). Consequently, data was pre-processed to include 689 chemical descriptors of the drugs and 138 genomic features for differentiating the cell lines, resulting in an input space of 827 features.

Chemical descriptors were generated with PaDEL software [19] from simplified molecular-input line entry system (SMILES) structures. Descriptors include physicochemical features such as weight, lipophilicity, rule of five, and additionally fingerprints of the drugs (for details see "Materials and Methods" section, "Features" subsection, and http://padel.nus.edu.sg/software/padeldescriptor/).

For building our model, we used GDSC screening data from 608 genomically characterised cell lines and 111 drugs for which chemical information were available (see Fig 2 and Methods for details). The published version of this matrix holds 38,930 $IC_{50}$ values (~58% of the total, due to technical and logistic reasons).

We performed an 8-fold cross-validation, where the test set of each fold was not used for training so as to measure the predictive power of the resulting models across all drugs rather than for each drug separately. Neural networks were able to impute missing $\log(IC_{50})$ values on the test sets with an averaged Pearson correlation coefficient ($R_p$), coefficient of determination ($R^2$) and root mean square error (RMSE) (Text S1) of 0.85, 0.72 and 0.83 across all 111 drugs, respectively (Fig 3A). Alternatively, random forests achieved comparable performances ($R_p$ of 0.85, $R^2$ of 0.72 and RMSE of 0.84; full details in supplementary materials). Furthermore, we conducted a blind test using 13,565 new experimental $IC_{50}$ values only received after training our models in order to verify our cross-validation results (drug-to-cell line matrix updated by ~18%, with these newly generated $IC_{50}$s exclusively used as the blind test set). The results on the blind test were almost as good as in the cross-validation, obtaining an $R_p$ of 0.79, $R^2$ of 0.64 and an RMSE of 0.97 (Fig S1, Text S2). The accuracy of the predictions encouraged us to train the networks with fewer $IC_{50}$ values. Remarkably, the predictive power of the models did not fall appreciably off in quality, even if the amount of training data was reduced to 20 % of the total (Fig 3B).

Using an analysis of variance (ANOVA) to identify drug-to-oncogene associations, we investigated how well the $IC_{50}$ values predicted for the test set using our model recapitulate associations manifested in the experimental data, for instance, whether a given mutation is causing sensitivity or resistance against a drug [3]. Using only predicted $IC_{50}$ values, we correctly captured 79% (168/213) of the significant observations with the same t-test tendency (positive or negative effect on drug sensitivity) identified with the experimental $IC_{50}$s. When only considering significant associations from our model (p-value adjusted with Benjamini-Hochberg, FDR=0.2), we correctly predicted 28% (59/213) of all experimentally identified associations. Where we failed to detect an association the ANOVA effect size is often small, or the experimental correlation is associated with a mutation either not or infrequently represented within the subset of cell lines with predicted $IC_{50}$ values. Notably, as example of the utility of this approach, using only predicted $IC_{50}$ values we identified known drug-to-oncogene associations such as sensitivity of *BRAF*-mutated cells lines to *MEK1/2*-inhibitors (Fig 4B) [20]. The range of predicted $IC_{50}$ values for a drug are typically narrower than for the observed values and is likely because currently available genomic dataset are in sufficient to explain the observed range of drug responses across the cell lines.

In addition, we assessed the predictive power of our model for unknown cell lines. Therefore, we applied a more stringent 8-fold cross-validation, where a cell line was either included in the train or test set. These models achieved an $R_p$ of 0.82, $R^2$ of 0.68 and an RMSE of 0.89 (Fig S2) demonstrating the accuracy of our model to predict $IC_{50}$ values for completely new cell lines. In an additional simulation, we left out all cancer cell lines from a specific tissue, e.g. we removed all lung cancer cell lines (106 out of 608 cell lines) and still obtained an $R_p$ of 0.79, $R^2$ of 0.61 and RMSE of 0.99 (Fig S3).

**Discussion**

Our results show that by using genomic features from the cell lines and chemical information from drugs, it is possible to build *in silico* multi-drug models to impute missing $IC_{50}$ values with non-parametric machine learning algorithms such as neural networks and random forests. As output for our method, we chose to explore IC50 values as generated by Garnett et al. [3], which enables us to compare our results to them, however other metrics (such as a capped IC50 or area under the curve), might provide additional insight and potentially lead to more robust models.

The Pearson correlation (Fig. 2A) and coefficient of determination (Fig. 2B) of the multi-drug model are significantly better than the single-drug models, while the RMSE error is similar (Fig 2C). This means that the error (on average) of predicting a given IC50 value is the same in the multi-drug and single-drug models (RMSE) and, since some drugs are active at different concentration ranges, the model is able to cover a much larger dynamic range with a similar precision. The coefficient of determination balances these two terms, and thus a broader range with the same RMSE increases $R^2$. Thanks to the use of chemical descriptors, multi-drug models are trained with a volume of data that is two orders of magnitude bigger than the data to train each single-drug model. This larger dataset weights the difficulty in training heterogeneous response values across drugs.

In several instances, the use of multi-drug models permitted the *in silico* identification of genomic events associated with altered drug sensitivity, which is only possible when genomic properties are considered. Although our models did not capture all known gene to drug associations, we anticipate that as larger drug sensitivity and genomic datasets become available in coming years the predictive power of these models will increase. We believe that the predictive power of our models is due to the large number of cell lines and broad range of drugs in the GDSC panel that samples intensively the chemical space of common cancer drugs (chemotherapeutic and kinase inhibitors). It remains to be determined how these models will predict completely unknown families of therapeutic agents.

The predictive ability of our methods for individual values is still limited and could be further improved by extending the set of input features with additional layers of molecular characterization of the cell lines, such as basal transcriptional profiles and phosphoproteomic data. These data types have been used to predict drug responses in various contexts [21-24]. Another valuable extension could be the inclusion of gene expression data following drug treatment, a powerful *in silico* resource for predicting treatment outcomes and elucidating compound mode of action [25,26], as well as a promising gateway to the identification of new drug repositioning opportunities [27]. Additionally, epigenetics data could enhance the prediction capabilities of future methods [28].

Our method uses purely experimental data, but additional predictive power can be expected from including knowledge of the underlying network [29]. It has been shown that the prediction of drug response and mode of action by transcriptional profiling is significantly enhanced when paired with known a priori gene and protein networks [30,31] and drug similarities have been inferred based on the corresponding *in silico* predicted impinged pathway [32]. Prior knowledge could also increase the interpretability of the results. Known regulatory relationships between genes and transcriptional data [33] and protein networks [34] can be used to identify deregulated pathways, and be further linked to the genomic alterations that drive them [35], highlighting subnetworks of importance for drug response. Incorporation of these additional features will require a scheme to prioritize the input features based on their impact on the final trained model. Associations between features and outcomes could be explicitly unveiled by integrating in our models feature selections criteria and dimensionality reduction techniques. In terms of predictive models, we have used standard machine learning methods (neural networks and random forests), given their flexibility and robustness as predictive models. A fertile ground for further research is investigating the application of other modeling techniques, including linear regression methods (e.g. LASSO, ElasticNets).

Our results also show that one can estimate the accuracy of prediction for different degrees of sparseness in the data, which may have utility when designing experiments where coverage has to be balanced with accuracy. Furthermore, because models are able to predict $IC_{50}$ on cell lines not screened yet, predictions from these models can be used to decide whether it is worthwhile expanding the panel of cell lines, or rather focus on a few selected ones.

The implications of our results go beyond their utility to optimise the experimental design of drug screenings. Once a model is built, it could be used to systematically test the potential effect of novel drugs *in silico*, based on their chemical features and similarity. These predictions can help to evaluate the potential activity of new drugs, e.g. from large chemical libraries, to be screened. Furthermore, predictions on clinically approved drugs is expected to reveal candidates for drug repurposing and potentially identify specific disease sub-types that would be most responsive [8]. Although cell lines are not an exact replica of real tumours, comprehensive predictive models such as ours together with expanded genomic and epigenomic datasets may be a good proxy to facilitate the development new therapeutic strategies tailored to individual patients [12].

**Materials and Methods**

*Training dataset* - We used the data from the Genomics of Drug Sensitivity in Cancer project [3], which contains 639 cancer cell lines, each of them characterised by a set of genomic features (details in the next section). The characterisation is not complete for every cell line, and therefore we filtered out cell lines with more than 15 missing genomic features, which reduced the set of selected cell lines from 639 to 608. The dataset contains 131 drugs. As our method exploits the chemical structure of each drug, this information in simplified molecular-input line entry system (SMILES) format is required. Therefore, we did not consider the 20 drugs for which SMILES were not available, and built our model for the remaining 111 drugs.

The resulting matrix of 608 cell lines by 111 drugs will have 67,488 possible drug response curves, each summarised by its $IC_{50}$ value (drug concentration in μM units required to eradicate 50% of the cancer cells). Currently, the dataset contains 38,930 $IC_{50}$ values out of these 67,488 (58%), with missing values mostly due to logistic reasons such as co-ordinating measurements from various screening centres. The log $IC_{50}$ ranges from -7.40 ($IC_{50}$~$4·10^{-8}$ M; the most sensitive drug-cell combination) to 6.91 ($IC_{50}$~$8·10^{6}$ M; the most resistant). Note that extremely large and small values are extrapolations in the $IC_{50}$ that have no clinical relevance. We use these ranges in this study as those are the ones used in the paper Garnett et al. [3] that we compare our results against.

*Blind test dataset* - We generated test sets during the cross-validation for estimating the expected error (details in cross-validation section). However, even cross-validation can overestimate the prospective performance of machine learning methods. Therefore, we conducted a truly blind test in order to demonstrate the prospective capabilities of our cross-validated models to impute missing IC50 values in the 608 cell lines by 111 drugs matrix (Fig S1). Our blind test contains 13,565 newly generated $IC_{50}$ values, which were obtained after training took place, or put it differently, a batch of new experimental data was generated to independently validate our models. To sum up, 58% of the $IC_{50}$ values are in the original dataset (used for cross-validation), an additional 18% are used for the blind test (independent test).

*Features* - There are two different input data streams in our method: the genomic background for each cancer cell line, and the chemical properties of a drug. For the first input data stream, cancer cell lines are characterised by the mutational status of 77 oncogenes, where each of them is further described by copy number variation (any high grade amplification or homozygous deletion of a cancer gene) and sequence variation (changes in the protein sequence, e.g. non-synonymous single nucleotide polymorphism). Additionally, there is one binary feature for the microsatellite stability status of each cell line. The cell line features were encoded as followed:

$$\text{Microsatellite instability status} = \begin{cases} 1 & ,\textit{if}\quad \textit{unstable} \\ 0 & ,\textit{if}\quad \textit{stable} \end{cases}$$

$$\text{Sequence variation} = \begin{cases} 1 & ,\textit{if}\quad \textit{mutation} \\ 0 & ,\textit{if}\quad \textit{wildtype} \end{cases}$$

$$\text{Copy number variation} = \begin{cases} 1 & ,\textit{if}\quad \textit{amplification} \\ 0 & ,\textit{if}\quad \textit{wildtype} \\ -1 & ,\textit{if}\quad \textit{deletion} \end{cases}$$

All mutations considered, we have 77 possible copy number variations plus 77 possible sequence variations and one microsatellite stability value, which sums up to 155 possible cell line features. However, a few mutational features are missing for some cell lines, and we conservatively removed a feature in case it was missing for any cell line. This led to a final set of 138 genomic features characterising each cancer cell line.

The second input data stream incorporates 1D and 2D chemical properties of each drug. We generated these chemical features using the PaDEL software (v2.11, downloaded from the project website, http://padel.nus.edu.sg/software/padeldescriptor/) [19] from the SMILES with default settings. 722 features are physicochemical descriptors and 881 are obtained from the fingerprints, leading to a total of 1603 chemical features. We only included chemical features that could be calculated for all drugs. Furthermore, we removed any feature with the same value across all drugs, obtaining a final set of 689 chemical features for each drug (e.g. atom count, bond count, molecular weight, xlogP or PubChem fingerprint, to name a few). The list of drugs is available in the Supplementary material (druglist.csv). Taking together the cancer cell line and drug stream, we used 827 features to build our predictive models of the log $IC_{50}$ value of a given cell line in the presence of a given drug.

*Cross-validation* - We used an 8-fold cross-validation for building our models. Therefore, we separated the original dataset into eight equally sized sets of $IC_{50}$ values, obtained by randomly distributing all $IC_{50}$s of the matrix into 8 bins. One of them was exclusively used for testing (never involved in any training), other six were destined for training the model and the remaining piece was used for cross-training. Cross-validation is a process used to avoid under- and overfitting [36] e.g. identifying the optimal number of hidden units and training iterations for a neural network (details in "Machine learning" section). We rotated iteratively the sets so that each data point was used at least once for training, cross-training or testing. Finally, we obtained 8 models, which were equally predictive.

Furthermore, we used a more stringent version of the above described 8-fold cross-validation. We ensured that test, train and cross-train set are not sharing any cell line, which might occur in the non-stringent version (described above). For instance, assume cell line C1 is treated with the drugs D1, D2 and D3; For the non-stringent cross-validation, the combination C1-D1, C1-D2 and C1-D3 might be distributed over test, train and cross-train set; for the stringent cross-validation, every combination with C1 is exclusively occurring in one of those three sets.

*Machine learning* – For the neural networks, we used the Java implementation from Encog 3.0.1 (http://www.heatonresearch.com/encog) [37,38] of a feed-forward multi layer perceptron, where we defined three different layers: input, hidden (or middle) and output layer. Every perceptron of a layer is completely connected to each perceptron of the upper layer. The number of features determined the number of input units, or put it differently, required perceptrons in the first layer. The number of hidden units was explored during the training for determining the correct model complexity, which was between 1 and 30 hidden units. Furthermore, each input and hidden unit had also an bias, which is a permanent activation input for those perceptrons. We used a single output unit for predicting the continuous log($IC_{50}$) value.

As perceptron activation function for enabling the network to predict non-linear behaviour, we used the sigmoid function, which returns values in an interval from 0 to 1. Therefore, we had to normalise the $IC_{50}$ values (raw $IC_{50}$ values, not in log space) also into a range from 0 to 1, which was done with the following logistic-like function:

$$norm(y) = \frac{1}{1+y^{-0.1}} \quad where \quad y > 0$$

$y$: Observed/expected IC50 value, which has to be a positive number greater than zero.

We trained the network with the resilient error backpropagation implementation from Encog with default parameters [39]. For exploring the final model complexity, which is described by number of hidden units and amount of training iterations, we examined different neural network architectures from 1 up to 30 hidden units and trained them for maximal 400 iterations. We searched the global minimum in that cross-training landscape (minimizing the root mean square error of cross training set) for avoiding an under- or overfitting (usually, between 21 and 27 hidden units were chosen as best model after approximately 300 iterations).

We also carried out random forest [40] regression models to investigate whether there was any significant performance gain using an alternative non-parametric machine learning methodology (Text S3). A random forest is an ensemble of many different regression trees randomly generated from the same training data (recommended value of n=500 trees was used).

*Data access* - The dataset is fully accessible of the Genomics of Drug Sensitivity in Cancer project [3], downloaded from the project website, http://www.cancerrxgene.org/. The training set is based on release v1.0 from June 2012. Newly generated $IC_{50}$ values of the blind test are published in release v1.1 from July 2012, which are not part of Release v1.0.

*Software access* – The Encog Machine Learning Framework (version 3.0.1) [37,38] containing the neural network implementation is a free available and open source (Apache License 2.5), and could be downloaded on the Heaton Research webpage (http://www.heatonresearch.com/encog). For the random forest model, the R package randomForest (version 4.6-6) [41] is also freely available under GPL licence from CRAN webpage (http://cran.r-project.org/web/packages/randomForest/index.html).

**Acknowledgements**

We thank King Wai Lau, David Wedge and Jorge Soares for helping with data, Marc Hafner, Mario Niepel, John Marioni, Theo Knijnenburg, Lodewyk Wessels for useful discussions, and Clare Pacini and Maja Köhn for feedback on the manuscript.

**Figures:**

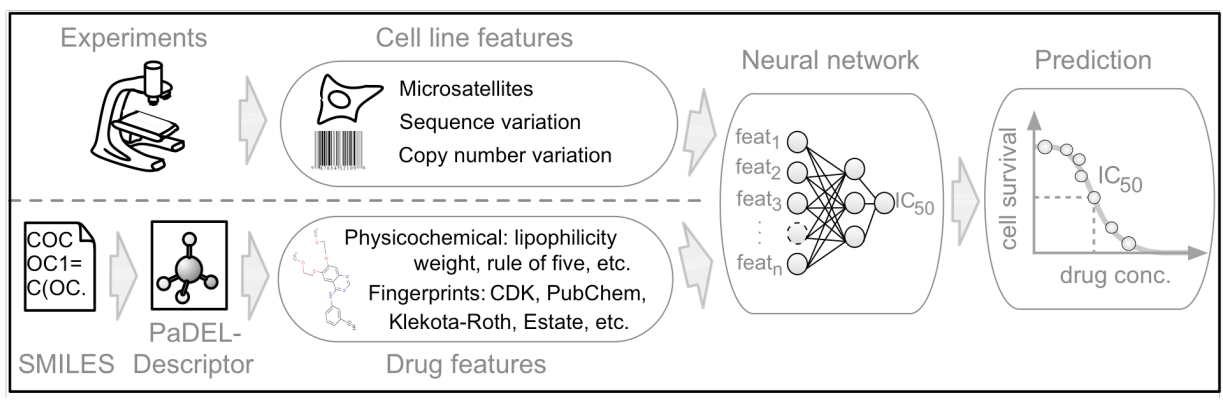

**Figure 1: IC$_{50}$ prediction workflow.** Our method is based on two different input streams: (1) cell line features of 77 oncogenes and their mutation state, (2) drug features that are generated with PaDEL software [19] from the simplified molecular-input line entry system (SMILES), see method section for details. The continuous IC$_{50}$ value is predicted with state-of-the-art machine learning algorithms (neural networks and random forests).

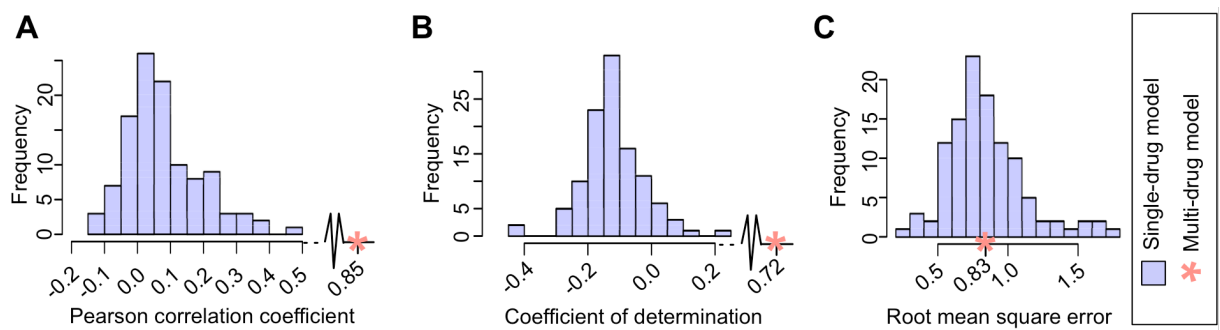

**Figure 2: Comparison of single-drug models and the multi-drug model.** The performance of the multi-drug model (red asterisk) and the family of 111 single-drug models (blue histogram) is represented using three different metrics: (A) Pearson correlation $R_p$, (B) coefficient of determination $R^2$, and (C) root mean square error RMSE.

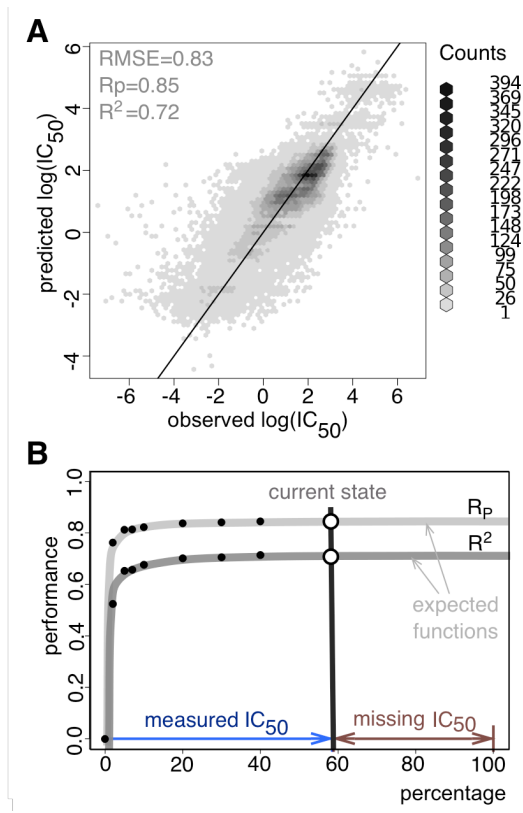

**Figure 3: IC$_{50}$ prediction.** Predictions are achieved with 8-fold cross-validations. Performance values are exclusively calculated on the test sets. (A) Correlation between predicted to experimental observed log(IC$_{50}$) values (Pearson correlation R$_p$ = 0.85 ; coefficient of determination R$^2$ = 0.72, root mean square error RMSE = 0.83). Although there is an enrichment of resistant cell lines, which tend to have higher log(IC$_{50}$) values than sensitive cell lines, the lower log(IC$_{50}$) values are still decently predicted. (B) Expected improvement of the IC$_{50}$ prediction by filling experimentally gaps in the cell-to-drug matrix. The vertical grey line corresponds to the published data set (filled to ~58%, due to logistic reasons), which corresponds to the results in panel (A). However, similar accuracies (R$_p$ of 0.84 instead of 0.85, R$^2$ of 0.70 instead of 0.72) can be achieved using exclusively 20 % of the whole matrix.

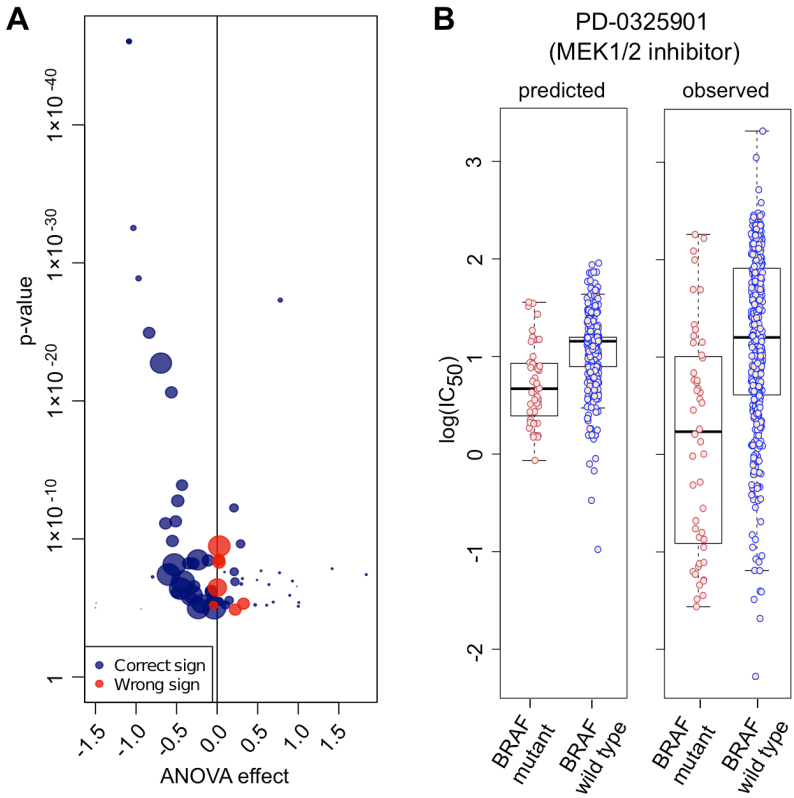

**Figure 4: Comparing ANOVA with prediction.** (A) Analysis of variance (ANOVA) of experimental data and predicted output for drug-to-oncogene associations (20% FDR). The size of each association (dot) is proportional to the amount of treated cell lines containing the particular mutated oncogene. Blue dots indicating the same t-test tendency in our predictions, and red ones the opposite. (B) Predicted and measured IC$_{50}$s of *BRAF*-mutated vs. wild-type cell lines exposed to the *MEK1/2*-inhibitor PD-0325901 (p-value of prediction = $1.91 \times 10^{-05}$, t-test multiple hypothesis corrected with Benjamini & Hochberg).

**Supplemental Materials**

**Text S1: Measuring performance.**

The output of our method is the logarithm to base 10 of the IC$_{50}$ in µM:

$$y = \log(IC_{50})$$

$y$: Prediction of log(IC$_{50}$) by our method

$IC_{50}$: Required concentration [µM] to eliminate 50% of the cancer cell population, computed as described in Garnett et al *Nature* 2012

We evaluated three different performance metrics, (i) one that captures the linear dependency of the prediction versus observation, (ii) a metric for outlining the variance of an assumed perfect prediction and (iii) additionally another metric that describes the average error of the model predictions:

(i) Pearson correlation coefficient ($R_p$) describes the relationship of prediction and observation. $R_p$ is in a range from -1 to 1; negative correlations hint at inverse predictions (more predicted wrong than correct), $R_p$ of 0 correspond to a random relationship (no correlation), and positive correlations indicate linear behaviour with positive gradient:

$$R_p = \frac{\sum \vec{y}_{obs} \vec{y}_{pred} - \frac{\sum \vec{y}_{obs} \sum \vec{y}_{pred}}{n}}{\sqrt{\left(\sum \vec{y}_{obs}^{\,2} - \frac{1}{n}\sum \vec{y}_{obs}^{\,2}\right)\left(\sum \vec{y}_{pred}^{\,2} - \frac{1}{n}\sum \vec{y}_{pred}^{\,2}\right)}}$$

$n$: Size of the test set

$\vec{y}_{pred}$: Vector of observed/expected log(IC$_{50}$) value

$\vec{y}_{obs}$: Vector of predicted log(IC$_{50}$) value

(ii) Coefficient of determination ($R^2$) measures the proportion of the variance of the data that is explained by the regression model. As regression model, we assume a linear function representing a perfect prediction, or put a differently, plotting a line through observation-by-observation points. The following definition of $R^2$ typically returns values in range from 0 to 1, where values closer to one indicate a good prediction and 0 suggest weak fitting of the observations. However, since the regression model is not data driven and rather a conservative assumption of being a perfect prediction, also negative values are possible in case the prediction is far off the observation.

$$R^2 = 1 - \frac{\sum_{i=1}^{n}(y_{obs_i} - y_{pred_i})^2}{\sum_{i=1}^{n}(y_{obs_i} - \bar{y}_{obs})^2}$$

$n$ :     Size of the test set

$y_{obs}$ :  Observed/expected log(IC$_{50}$) value

$\bar{y}_{obs}$ :  Average of all observed log(IC$_{50}$) values

$y_{pred}$ :  Predicted log(IC$_{50}$) value

(ii) Root mean square error (*RMSE)* provides an average of the error across all predictions made by the models:

$$RMSE = \sqrt{\frac{1}{n}\sum_{i=1}^{n}(y_{obs_i} - y_{pred_i})^2}$$

$n$ :     Size of the test set

$y_{obs}$ :  Observed/expected log(IC$_{50}$) value

$y_{pred}$ :  Predicted log(IC$_{50}$) value

**Text S2: Comparing performance of imputation methods and machine learning approach.** A straightforward way to estimate missing IC$_{50}$ values is to impute them from the rest of the values. We therefore compared our feature-based approach against conventional imputation methods on our blind test. We used well-known reference methods from the R package "imputation" v1.3 developed by Jeffrey Wong: Singular Value Threshold (SVT), Singular Value Decomposition (SVD), and k-Nearest Neighbor (kNN) imputation. To estimate the parameters of each imputation method, we used cross validation as suggested in the example code of the R package. SVT imputation obtained 0.06 $R_p$ and 1.66 *RMSE*. The SVD imputation improved the performance to 0.37 $R_p$ and 1.54 *RMSE*. The best performing imputation method is kNN with 0.71 and 1.10 $R_p$ and *RMSE*, respectively. Our method (in this case a neural network) including genomic and chemical drug properties outperformed all imputation methods with an $R_p$ of 0.79, $R^2$ of 0.72 and *RMSE* of 0.97. It is important to note that imputation techniques, unlike feature-based machine learning, cannot be used to extrapolate, i.e. to predict IC50s for unseen cell lines or drugs.

**Text S3: Random Forest.** In model building, we used the same training set, features and performance measures as in the neural network model. However, unlike in the neural network and for the sake of efficiency, we did not tune any parameter of the Random Forest and hence the 8-fold cross-validation was slightly different (i.e. the same test set used by the neural network model, but exploiting more training data with a control parameter-free Random Forest). Here all seven partitions are exclusively used for training and the last partition for test. As usual, the prediction on the eight independent folds, one per each regression, is averaged.

In the case of randomly constructed partitions, *RMSE* was 0.84, $R^2$ was 0.72 and the $R_p$ was 0.85. In the case of leaving cells out partitions, *RMSE* was 0.85, $R^2$ was 0.71 and the $R_p$ was 0.84. The performance on the blind test was *RMSE* was 1.00, $R^2$ was 0.59 and the $R_p$ was 0.78. Overall, these results are very similar to those obtained with the neural network model. Each machine learning model was independently trained and validated using different computer codes, which further supports the robustness of our results.

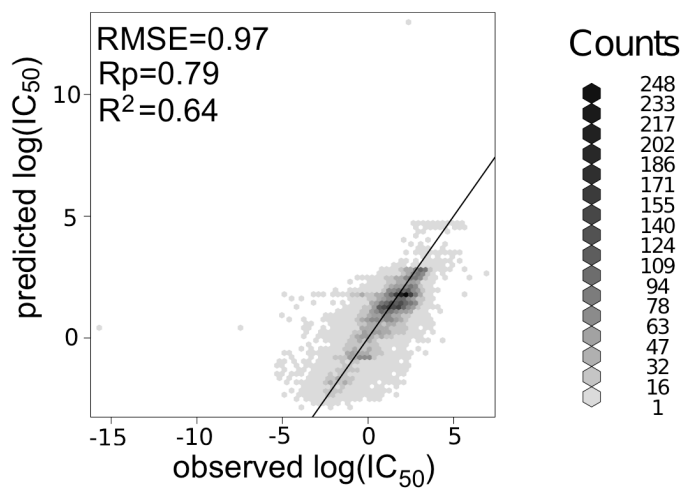

**Figure S1: Blind test of multi-drug model.** The training dataset holds 38,930 $IC_{50}$ values, that is ~58% of all possible drug-to-cell line combinations. For the blind test, 13,565 novel $IC_{50}$ values were generated, an~18% additional data points which were not included in the training dataset. For obtaining the predicted log(IC50) values, we averaged the output of each model (8 different models resulting from the 8-fold cross-validation procedure). The prediction on the blind test was slightly worse than that estimated by cross-validation (Fig 3A): root mean square error (*RMSE*) was increased from 0.83 to 0.97, coefficient of determination ($R^2$) declined from 0.72 to 0.64 and the Pearson correlation coefficient ($R_p$) was decreased from 0.85 to 0.79. This small performance decrease is due to the fact that blind test data points are not selected at random: these tend to come from drug-cell combinations that are not optimally represented in the training set (i.e. those cell lines in the training set that have been probed against every drug in the panel will not have further IC50 values in the test set, as all training and test sets in this study are non-overlapping).

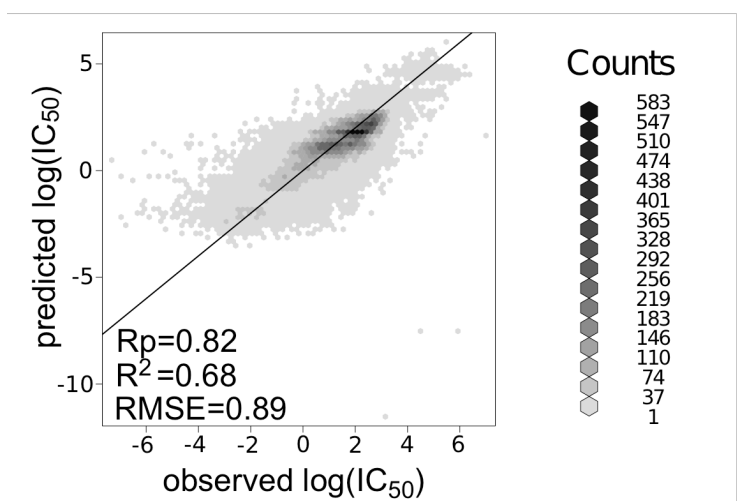

**Figure S2: Correlation between predicted to experimental observed log(IC$_{50}$) values leaving out cell lines.** The stringent 8-fold cross-validation was performed on the distinct set of cell lines, so that a cell line was neither used for testing or involved in the training. The figure shows values obtained solely on the test sets. The prediction quality is slightly worse than the normal cross-validation (Figure 3A): *RMSE* increased from 0.83 to 0.89, R$^2$ decreased from 0.72 to 0.68 and the $R_p$ decreased from 0.85 to 0.82.

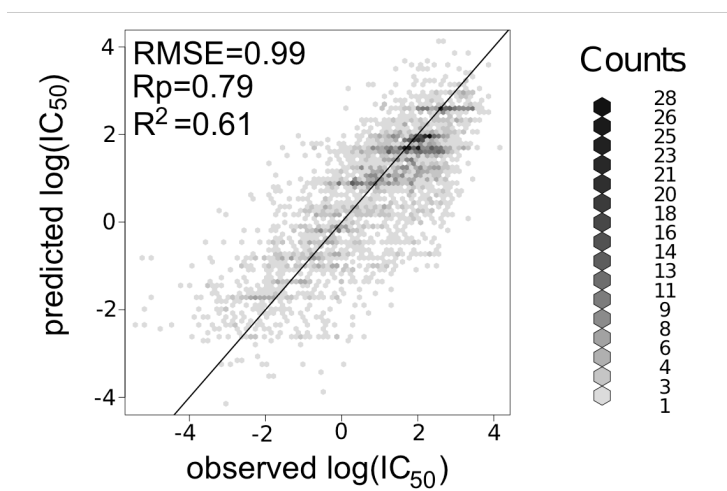

**Figure S3: Correlation between predicted to experimental observed log(IC$_{50}$) values leaving out all lung cell lines.** To further challenge our model and our hypothesis that it is possible to leave out several cell lines, we removed all lung cell lines and used them exclusively for testing. There are 106 out of 608

cell lines are from lung tissue (~17 % from data), which we were able to predict with minor performance reduction compared to including all cell lines (Figure 3A): root mean square error (*RMSE*) increased from 0.83 to 0.99, coefficient of determination ($R^2$) declined from 0.72 to 0.61 and the Pearson correlation coefficient ($R_p$) decreased from 0.85 to 0.79.